\documentclass[twocolumn,prb]{revtex4}
\usepackage{graphicx}
\begin{document}
\title{Variational perturbation approach to the Coulomb electron gas}
\author{Sang Koo You   and   Noboru Fukushima}
\affiliation{Max-Planck-Institut f\"{u}r Physik komplexer Systeme,
         N\"{o}thnitzer Strasse 38, 01187 Dresden, Germany}

\date{11 April 2002}
\begin{abstract}
The efficiency of the variational perturbation theory 
[Phys. Rev. C {\bf 62}, 045503 (2000)] formulated recently for
many-particle systems is examined by calculating
the ground state correlation energy of the 3D electron gas with the Coulomb 
interaction. 
The perturbation beyond a variational result can be carried out systematically
by the modified Wick's theorem which defines a contraction rule
about the renormalized perturbation.
Utilizing the theorem,  
variational ring diagrams of the electron gas are summed up.
As a result, the correlation energy is found to be much closer to the result of 
the Green's function Monte Carlo calculation
than that of the conventional ring approximation is.
  
\end{abstract}

\pacs{05.30.Fk, 71.10.Ca}
\maketitle
\section{Introduction}
\vspace*{-0.2cm}
A variational method has been widely used as a convenient and powerful tool 
to calculate physical quantities such as energy and order parameters [1-5].
The most important advantage of this approach consists in its applicability
even for the strongly correlated systems through suitable trial wavefunctions.
However, it is extremely difficult to improve the results for the systems with
many degrees of freedom.
In contrast, results of the perturbative approach are systematically improved by
calculating the further corrections using the Feynmann rule [6-8].
However, the convergency can not be guaranteed in the unweakly coupled cases.

There have appeared many efforts trying to combine 
the above two approaches in order to overcome each drawback [9-14].
Many studies have been performed mainly in the areas of relativistic 
field theories and quantum mechanics under the principle of minimal sensitivity (PMS) [9], 
which says that approximated physical quantities in the perturbation theory 
should be minimized with respect to the parameters absent in the original Hamiltonian.  
These studies are called optimized perturbation theory or variational perturbation
theory (VPT), which is applied very successfully 
to  quantum mechanical problems such as the anharmonic oscillator and the double-well potential.
However, we are faced with  extremely complex higher-order
calculations in the case of quantum field theories.

Recently, papers on another kind of VPT have been published [15-18].
This approach is not related to the PMS, 
because the minimization is carried out at the zeroth order
and the variational parameters are fixed before the perturbative calculation. 
It has a simple expansion rule, which
makes the VPT approach to many-particle systems more manageable.  
In this paper, the VPT for fermion systems of ref. [17] is briefly reviewed, and 
as a test on the efficiency of the method, we calculate the ground state
correlation energy of the three dimensional Coulomb electron gas  
within a ring approximation.
\vspace*{-0.5cm}

\section{Variational perturbation theory}
\vspace*{-0.2cm}
In this section, we briefly review the VPT 
in the functional integral formalism [17]. 
The partition function for fermion systems with two-body interaction
is generally given by
\begin{equation}
Z = \int D[\bar{\psi}\psi] \ \exp({-G_{ij}^{o^{-1}}
 \bar{\psi}_{i} \psi_{j} - V_{ijkl}
\bar{\psi}_{i} \bar{\psi}_{j} \psi_{k} \psi_{l} }) \ ,
\end{equation}
where $\bar{\psi}_i$,$\psi_i$ are Grassmann variables and 
each subscript stands for all possible quantum numbers 
including an imaginary time.
The  integration and summation symbols are omitted by the summation convention.
Here, $G^o $ is the bare Green's function matrix and $V$ is an interaction tensor.
The second term of the exponent in Eq.(1) is the perturbative term in the conventional
perturbation theory.
Now, introducing a variational Green's function $G$, the exponent is 
rewritten as
\begin{eqnarray}
-G_{ij}^{-1}
\bar{\psi}_{i} \psi_{j}-( G_{ij}^{o^{-1}} - G_{ij}^{-1} )
 \bar{\psi}_{i} \psi_{j} -V_{ijkl}
\bar{\psi}_{i} \bar{\psi}_{j} \psi_{k} \psi_{l}  \ .
\end{eqnarray}
Then, the second and third terms in Eq.(2) are regarded as a renormalized interaction.
According to the Jensen-Peierls' inequality [19,20], we get the inequality between
the thermodynamic potential $\Omega$ and the variational thermodynamic potential 
$\bar{\Omega}(G)$; $\Omega \leq \bar{\Omega}(G)$. 
Minimizing $\bar{\Omega}(G)$ with respect to $G$, we obtain an equation of $G$ as 
\begin{equation}
G^{-1} - G^{o^{-1}} + \Sigma = 0 \ ,
\end{equation}
where $\left< b| \Sigma |a \right> \equiv 2 ( V_{ibaj} - V_{biaj}) G_{ji}$.
We note that
$G$ corresponds to the self-consistent Hartree-Fock Green's function for the
interacting fermion system.
Using the condition Eq.(3), 
the minimized variational thermodynamic potential $\bar{\Omega}_{\rm min}$ 
and the correction part $\Delta \Omega \equiv \Omega - \bar{\Omega}_{\rm min}$ 
 are arranged as follows,
according to the notation in ref.[17];
\begin{eqnarray}
 && \bar{\Omega}_{\rm min} = \frac{1}{\beta}{\rm Tr} \ln G - 
  \frac{1}{2 \beta}{\rm Tr} \Sigma G  \ ,  \\
 && \Delta \Omega =
-\frac{1}{\beta} \left< \exp \left[-V_{ijkl} \left(  \partial_{ \eta_{i}}
 \partial_{ \eta_{j}}  \partial_{ \bar{\eta}_{k}}
 \partial_{ \bar{\eta}_{l} }  \right)'\right] \right>_{G, C}  \ ,
\end{eqnarray}
where $\left< \cdot \cdot \cdot \right>_{G} \equiv \cdot \cdot \cdot 
e^{ \bar{\eta}_{i} G_{ij} \eta_{j}} |_{\bar{\eta}, \eta = 0}$  represents the thermal
average using $G$
and the subscript $C$ indicates the connected contractions among all possible contractions
by the {\it modified Wick's theorem} Eq.(7) below.
During the derivation of the above equations, 
 Grassmann variables $\bar{\eta}$, $\eta$ are introduced 
as source fields of the original fields $\psi$, $\bar{\psi}$, and the renormalized interaction 
of Eq.(2) is expressed as functional derivatives about source fields.
After integrating out the Gaussian integral ($\int D[\bar{\psi}\psi]  \exp(- \bar{\psi}_{i}
G_{ij}^{-1} \psi_{j} + \bar{\eta}_{i} \psi_{i} +
\bar{\psi}_{i} \eta_{i})  = {\rm Det} G^{-1} e^{ \bar{\eta}_{i}
G_{ij} \eta_{j}}$) and rearranging the thermodynamic potential 
under the condition Eq.(3), we obtain  Eqs.(4) and (5). 
Here, the primed derivative operation $( \cdot \cdot \cdot )'$ is defined as
\begin{eqnarray}
&&\left( \partial_{ \eta_{i}}
\partial_{ \eta_{j}}  \partial_{ \bar{\eta}_{k}}
  \partial_{ \bar{\eta}_{l} }  \right)'     \nonumber \\
&&\equiv \left(  \partial_{ \eta_{i}}
 \partial_{ \eta_{j}}  \partial_{ \bar{\eta}_{k}}
 \partial_{ \bar{\eta}_{l} } - G_{li}  \partial_{ \eta_{j}}
 \partial_{ \bar{\eta}_{k}} 
 -G_{kj}  \partial_{ \eta_{i}}  \partial_{ \bar{\eta}_{l}}  
\right. \nonumber \\
 && \ \ \ + \left. G_{lj} \partial_{ \eta_{i}}  \partial_{ \bar{\eta}_{k}}
+G_{ki}  \partial_{ \eta_{j}} \partial_{ \bar{\eta}_{l}}
+G_{li} G_{kj} - G_{lj} G_{ki} \right) \ .
\end{eqnarray}
Carrying out the {\it n}th order calculation of the above operation,
we find {\it modified Wick's theorem } for the perturbative expansion 
of the present VPT
as
\begin{eqnarray}
&&\left<   \left( \partial_{ \eta_{i_1 }}
\partial_{ \eta_{j_1 }} \partial_{ \bar{\eta}_{k_1 }}
\partial_{ \bar{\eta}_{l_1 } } \right)' \cdot \cdot \cdot
 \left( \partial_{ \eta_{i_n }}
\partial_{ \eta_{j_n }} \partial_{ \bar{\eta}_{k_n }}
\partial_{ \bar{\eta}_{l_n } }  \right)'   \right>_{G}  \nonumber \\
&&={\rm Sum \ of \ all \ possible \ contractions } \nonumber \\
&& \ \ {\rm \ between \ different \ cells}. 
\end{eqnarray}
Here we call one $(\cdot\cdot\cdot)'$
a unit {\it cell} and the contraction rule is as follows;
each $\partial_{ \eta_{i}}$ is paired with a $\partial_{ \bar{\eta}_{j}}$ in
{\it a different cell} and then, each pair of derivatives
 $ \partial_{ \eta_{i}} \partial_{ \bar{\eta}_{j}}$
is replaced by Green's function $G_{ji}$. 
In this pairing, every permutation of two derivatives changes the sign.
The differences from the original { \it Wick's theorem} are to forbid the intracell contraction
and to use the renormalized Green's function $G$.
The unit cell with an interaction, 
$V_{ijkl}\left(  \partial_{ \eta_{i}}
 \partial_{ \eta_{j}}  \partial_{ \bar{\eta}_{k}}
 \partial_{ \bar{\eta}_{l} }   \right)' $, can be described by a diagram
as in Fig.1 (a). A contraction is to join two incomming and outgoing lines together and
to assign it a propagator $G$ which corresponds to a double line.
In Fig.1 (b), we show ring diagrams with renormalized propagators,  
where the first order diagram does not exist because it is an intracell-joining diagram.

 For example, 
 the second order perturbation of $\Delta \Omega$ is given by;
\begin{eqnarray}
&&-\frac{1}{\beta}  V_{ijkl} V_{i'j'k'l'} G_{l'i}G_{k'j}G_{kj'}G_{li'}
\nonumber \\
&&+\frac{1}{\beta}  V_{ijkl} V_{i'j'k'l'} G_{l'i}G_{k'j}G_{ki'}G_{lj'} \ .
\end{eqnarray}

\begin{figure}
\vspace*{-5.5cm}
\hspace*{-2cm}
\includegraphics[scale=0.6]{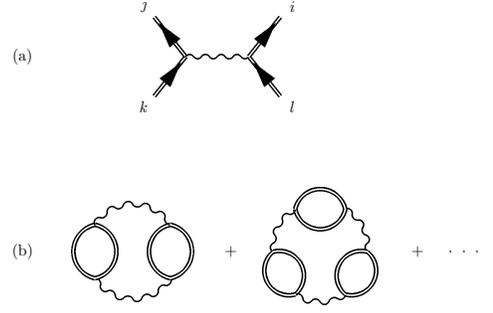}
\vspace*{-7.5cm}
\caption{(a) Unit cell to construct Feynmann diagrams; (b) the summation of 
         variational ring diagrams.
         the first order diagram cannot exist in the present theory. }
\end{figure}
\section{Coulomb Electron Gas}
The present formulation can be easily applied to any fermion systems with
two-body interaction.
In this section, the ground state correlation energy of the 
three dimensional Coulomb electron gas is 
calculated using the variational ring approximation which is described by ring diagrams
in Fig.1 (b). 
The model Hamiltonian is given by
\begin{equation}
H = \sum_{k \sigma} \epsilon_k  a_{k \sigma}^{\dagger} a_{k \sigma} + \frac{1}{2}
    \sum_{k k' q \sigma \sigma'} \hspace{-0.25cm} ^{'}  v(q) a_{k \sigma}^{\dagger} a_{k' \sigma'}^{\dagger}
    a_{k'-q \sigma'} a_{k+q \sigma} \ ,
\end{equation}
where $a_{k \sigma}^{\dagger}$($a_{k \sigma}$) is a creation(annihilation) operator
of an electron with wavevector $k$ and spin $\sigma$. We have defined
$\epsilon_k = \hbar^2 k^2 / 2 m_e $ and $v(q) = 4 \pi e^2 / V q^2 $. Here, $V$, $m_e$ and 
$e$ are the system volume, the electron mass and charge respectively.  The primed summation
indicates that the $q=0$ term is excluded because of the cancellation with the positively
charged background.
In this model, the interaction $V_{ijkl}$ of Eq.(1)  
is expressed as
\begin{eqnarray}
 V_{ijkl} &\rightarrow & V_{ k_{1} \sigma_{1} \tau_{1}, k_{2} \sigma_{2} \tau_{2},
 k_{3} \sigma_{3} \tau_{3}, k_{4} \sigma_{4} \tau_{4}} \nonumber \\
&=& \frac{1}{2} v( | k_4 - k_1 | ) \delta_{k_1 + k_2 , k_3 + k_4} \delta_{\sigma_1
\sigma_4}
\delta_{\sigma_2 \sigma_3}  \nonumber \\
&\times&
 \delta(\tau_1 - \tau_2) \delta(\tau_3 - \tau_4)
\delta(\tau_2 - \tau_3) \ .
\end{eqnarray}
The variational Green's function $G$ is determined by Eq.(3).
For a homogeneous system, $G_{ k \sigma \tau ,  k \sigma \tau'}$ is written as
$G_{k \sigma}(\tau -\tau')$ and the Fourier transform of Eq.(3) with respect to time gives
\begin{eqnarray}
G_{k \sigma}(i \omega_n ) =-\frac{1}{i \omega_n - 
( \tilde{\epsilon}_{k \sigma}  - \mu )} \ ,
\end{eqnarray}
where  
$\omega_n ={(2n+1)\pi}/{\beta}$ is the fermion Matsubara frequency with integer $n$, and
$\tilde{\epsilon}_{k \sigma }=\epsilon_k  - \sum_{q} v(q) n_{k+q \sigma}$ 
is the renormalized electron energy  using
 the Fermi distribution function with the renormalized energy,
$n_{k \sigma}={1}/{(e^{\beta(\tilde{\epsilon}_{k \sigma}-\mu)}+1)}$.

Therefore, the minimized variational thermodynamic potential $\bar{\Omega}_{\rm min}$ 
at $T = 0 $K is
\begin{eqnarray}
  \bar{\Omega}_{\rm min} 
 = \sum_{k \sigma} (\tilde{\epsilon}_k -\mu) n_{k \sigma}
  + \frac{1}{2} \sum_{k q \sigma} v(q) n_{k+q \sigma} n_{k \sigma} \ .
\end{eqnarray}
This is the Hartree-Fock result. The second term cancels
 the doubly counted interaction energy in the first term to
give the singly counted result. 
If the interaction part of $\tilde{\epsilon}_{k \sigma}$
and the second term are summed up, $\bar{\Omega}_{\rm min}$ has the same form as 
the conventional first order result $\Omega_1 $; 
$\Omega_{1}=\sum_{k \sigma} ({\epsilon}_k -\mu) n^{0}_{k \sigma}
  - \frac{1}{2} \sum_{k q \sigma} v(q) n^{0}_{k+q \sigma} n^{0}_{k \sigma}$ ,
 where $n^{0}_{k \sigma}={1}/{(e^{\beta({\epsilon}_{k \sigma}-\mu)}+1)}$. 
In addition, the Fermi distribution function with a renormalized energy 
is a step function in the ground state like that with a bare energy. 
Therefore, the Hartree-Fock ground state energy is equal to the conventional first order one,
which results from the spherical symmetry of electron gas [21].
Hence, one might think that
the ground state energy obtained by higher order calculations through 
the present VPT would produce the same result as that of the conventional perturbation method.
However, as we will see below, they are different because 
the VPT propagator has a renormalized energy $\tilde{\epsilon}_{k \sigma}$
and furthermore, 
there are some forbidden diagrams in the VPT expansion, which
are allowed in the conventional perturbation.

Since the present formalism parallels exactly the conventional perturbation except for
the modification of the Wick's theorem and the renormalization of the propagator,
we can carry out the ring diagram summation without any difficulty.
The ring contribution to the thermodynamic potential is depicted by Fig.1 (b).
The double line represents a renormalized propagator $G$. 
The variational
 ring diagrams are summed up to the logarithmic function as in the conventional ring 
 approximation [6], namely,
\begin{eqnarray}
\frac{1}{2 \beta} \sum_{q \nu_n} \ln [1+v(q)F(q, \nu_n )]
-\frac{1}{2 \beta} \sum_{q \nu_n} v(q)F(q, \nu_n ) \ ,
\end{eqnarray}  
where $\nu_n =2 \pi n / \beta$ is the boson Matsubara frequency and $F(q, \nu_n )$
is the renormalized Lindhard function,
\begin{eqnarray}
F(q, \nu_n ) = \sum_{k \sigma} \frac{n_{k+q, \sigma}-n_{k, \sigma}}
{i \nu_n -(\tilde{\epsilon}_{k+q, \sigma}-\tilde{\epsilon}_{k, \sigma})} \ .
\end{eqnarray}

At zero temperature, the discrete frequency summation $\frac{1}{2 \beta} \sum_{\nu_n}$ 
is replaced by a continuous integral $\frac{1}{4 \pi} \int d\nu$.

Figure 2 shows Lindhard functions at zero temperature. 
The small(large) dotted line corresponds to the bare(renormalized)
Lindhard function.
According to the linear response theory, the Lindhard function $F_0 (q, \nu)$ 
has important properties that it
is proportional to 
the spin and charge susceptibilities of free fermion systems,
$\chi_0^s (q, \nu)$ and $\chi_0^e (q, \nu)$. In addition, $F_0 (0, 0)$ is equal to the 
density of states at the Fermi surface, $D(\epsilon_F)$ [22]. We note that the 
 values of $F(q, 0)$ is 
much reduced by the inclusion of the exchange energy into a bare energy band.  
Furthermore, the behavior near $q=0$ is completely different,
which originate from
the much larger energy slope around the Fermi surface than the bare one. 
The difference between $F (q, \nu)$ and $F_0 (q, \nu)$ results in different 
ground state energies
between the variational and conventional ring approximations.

\begin{figure}
\vspace*{-4cm}
\hspace*{-0.7cm}
\includegraphics[scale=0.47]{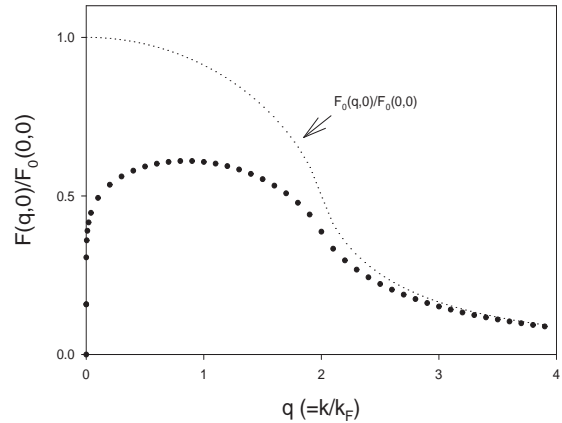}
\vspace*{-4.5cm}
\caption{The variational Lindhard function $F(q, 0)$(large dotted line) 
          at the zero temperature and frequency; the small dotted line represents the
          free electron Lindhard function $F_0 (q, 0)$. The magnitudes are renormalized
          by $F_0 (0, 0)$. }
\end{figure}

The chemical potential $\mu$ is an independent variable of the thermodynamic potential
$\Omega(\mu)$, hence in order to obtain the free energy as a function of electron density, 
we should change the independent variable using the Legendere transformation;
$F(N) = \Omega(\mu) + \mu N$ and $-\frac{\partial \Omega}{\partial \mu} = N$, where
$N$ is the number of particles and $F(N)$ is the free energy. However, this relation can
not be applied directly in approximations except for Hartree-Fock's and 
 Baym's self-consistent schemes [23] because 
$-\frac{\partial \Omega}{\partial \mu}$ does not represent the particle number.
Instead, we use the fact that the Fermi wavevector does not change 
with the inclusion of correlation according to Luttinger's theorem [24].
Therefore, we approximate the free energy as follows;
$F(N) \simeq \bar{\Omega}_{\rm min}(\mu_0 ) + \Delta \Omega_{\rm ring}(\mu_0 )
+ \mu_0 N $ and $-\frac{\partial \bar{\Omega}_{\rm min}}{\partial \mu} |_{\mu_0} = N$,
where the Fermi wavevector (or the density) is fixed by the second relation without
the correlation part. In the case of the conventional ring approximation, 
if $\bar{\Omega}_{\rm min}$ is replaced by a non-interacting part $\Omega_0 $,
the free energy by this transformation reproduces
the RPA(random phase approximation) correlation energy. 

Figure 3 shows the 
correlation energy as a function of $r_s$ which is the averaged relative distance between 
electrons defined by $V= (4/3)\pi (a_0 r_s)^3 N$, where $V$, $N$ and $a_0$ are 
the  system volume,  the number of electrons and the Bohr radius, respectively.
The triangles are the result
of the Green's function Monte Carlo calculation [25] and 
the better agreement with it is realized by the VPT. 
The conventional rings contain only direct Coulomb interacting processes
without exchange contributions which are very
important for fermion systems, and thus, the summation of them
 gives, so called, the RPA result [26]. On the other hand,
 the variational rings possess infinitely many higher order exchange
processes through the variational propagator $G$ as well as direct processes. 
As a result, we achieve much
improvement in the correlation energy beyond the RPA.  It is well known that the 
ring approximation can be applied satisfactorily for very small $r_s$ ($r_s \ll 1$).
In Fig.3, however, we note that even for large $r_s$, the variational rings give 
successful results. 

\begin{figure}
\vspace*{-4cm}
\hspace*{-0.7cm}
\includegraphics[scale=0.47]{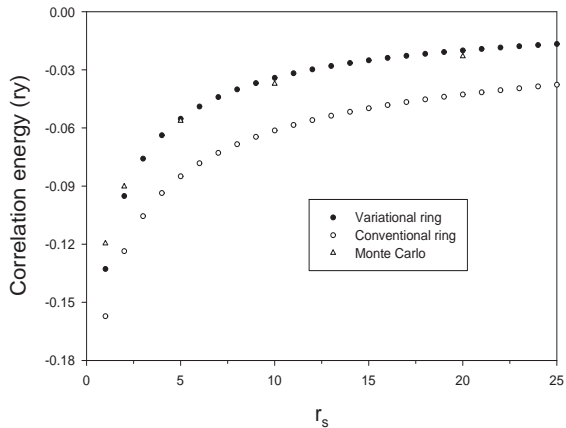}
\vspace*{-4.5cm}
\caption{The ground state correlation energy per particle calculated with ring diagrams;
         the solid and open circles are results of the variational and 
         conventional ring approximations, respectively. The triangles represent the result
         of the Green's function Monte Carlo calculation.  }
\end{figure}

Here, we examined the efficiency of the VPT by calculating the correlation energy of the Coulomb
electron gas. Further corrections beyond variational rings and the extension to 
finite temperatures are also easily accessible within the present VPT and we expect 
good applications to any other systems.

\section{summary}
After briefly reviewing  the VPT formulated recently 
for the many-particle systems, we have applied this method to the Coulomb electron gas.
The ground state correlation
energy of the Coulomb electron gas is calculated with the ring approximation of
the VPT. Improvement is achieved by the expansion with
 variational propagators. Especially, we note that
 even for large $r_s$, the approximation produces
quite successful results, as is evident from the good agreement with 
that of the Green's function Monte Carlo calculation.

\acknowledgments
This work was supported by postdoctoral fellowships program from Korea Science 
\& Engineering Foundation (KOSEF), and it was also financed in part by the Visitor
Program of the MPI-PKS.

\end{document}